# GENERATING EPICS IOC DATABASES FROM A RELATIONAL DATABASE – A DIFFERENT APPROACH

R. Keitel, TRIUMF, Vancouver, B.C., V6T 2A3, Canada

Abstract

The EPICS based control system of the ISAC radioactive beam facility uses the Capfast schematic editor to construct the IOC function block databases. This allows a self-documenting graphical representation of the IOC software using a hierarchical, object-like software structure with reusable components. On the other hand, the control system is quarterbacked by a relational database which contains all device information. Using database reports and Perl scripts, a device instantiation method was developed which generates top-level schematic files in Capfast format. This method exploits the advantages of device data entry and reporting using a relational database system while maintaining the graphical representation of the IOC database.

## 1 INTRODUCTION

One of the powerful features of the EPICS control system tool-kit is the method with which the software functionality of the front-end Input-Output Computer (IOC) is generated. Instead of writing conventional code, a run-time database is configured from standard function blocks.

In "EPICS-speak", these function blocks are called "records", and the collection of all records in an IOC is called "database". Although these terms are badly chosen, because they suggest the passive behaviour of a relational database, they will be used in this paper in the forms "EPICS record" and "IOC database" or "runtime database" to distinguish them from their relational counterparts.

## 2 EPICS DATABASE GENERATION

Any IOC database generation starts with assigning field values to EPICS records with the help of an IOC database configuration tool. Field values include hardware addresses, execution links to other EPICS records, and execution information such as membership in periodic scan groups.

Completely configured runtime databases are loaded into the IOC at startup. The EPICS runtime system (iocCore) establishes the required data structures and resolves the links between EPICS records within the same and between different IOCs.

As the IOC database files are ASCII files, a text editor is in principle sufficient to configure simple databases. More sophisticated, interactive tools add convenience to this process. With increasing complexity of a system, manual maintenance of the IOC database degenerates from cumbersome to error-prone to impossible. It is obvious to try and harness the power of a relational database system (RDB) for this process.

Different approaches have been taken, depending on the design of the underlying IOC database. If the IOC database consists of only a few EPICS records per device, it is possible to generate the IOC database files directly as reports from the RDB. If the device functionality expressed by EPICS records is more complex, EPICS database *templates* can be generated manually with a database configuration tool and instance parameters from an RDB can be inserted by some merging tool [1,2] or the whole process can be done by a more sophisticated RDB application [3].

## 3 EPICS DATABASES AT ISAC

For the ISAC project, the CAPFAST schematics editor [4] was chosen as the IOC database configuration tool. It provides a graphical representation of the EPICS records and their interactions and – most importantly – allows the implementation of a schematics hierarchy. This is done by introducing "symbols", each of which represents, but hides, the detailed functionality of a corresponding schematic. The symbol contains instance macros and hierarchical connections. Symbol instances are then used at the next higher level of the schematic hierarchy. Top-level schematics are translated into IOC databases using standard CAPFAST/EPICS tools. During this translation process, the schematic hierarchy is flattened.

Using CAPFAST, two types of IOC databases are configured at ISAC:

a)  Sub-system databases which contain all device instantiations
b)  "Other" IOC databases, which contain no device instantiations.

This paper is only concerned with sub-system databases.

## 3.1 Hierarchy

ISAC devices are implemented in an "object-like" way, using the hierarchical capabilities of the CAPFAST tool. Device behaviour, which is common to more than one device class, is implemented in "component" schematics. The components are abstracted and represented by component symbols. This leads to the following schematic hierarchy in the ISAC control system:

- Sub-system schematics – contain device instantiations (device symbols). They are not abstracted and are translated into an IOC database which is loaded into an IOC
- Device schematics – contain component symbols and may contain EPICS record symbols. They are abstracted by a device symbol.
- Component schematics – they may contain other component symbols and may contain EPICS record symbols. They are abstracted by a component symbol.
- Primitive component schematics – they contain only EPICS record symbols. They are abstracted by a component symbol.

Initially at ISAC, all sub-system schematics were configured interactively.

## 4 AUTOMATIC EPICS DATABASE GENERATION

In moving to automatically generated IOC databases, it was deemed very desirable to maintain the graphic representation of the sub-system databases, so that all run-time databases, which are loaded into IOCs, can be visually inspected the same way.

### 4.1 ISAC Relational Device Database

All ISAC device information is maintained in a relational database. For historical reasons, this device database is implemented using the Paradox RDB system. The device database is organized around the concept of a "logical device", which represents all "physical devices" of an operational unit. Examples of logical devices are:

- A magnet, consisting of the physical devices coils, power supply, water switch, and temperature switch
- A gate valve, consisting of the physical devices solenoid valve and limit switches

Each logical device is a member of one and only one functionality class. All members of a functionality class share the same behaviour and are distinct by their instance parameters, such as device name, hardware type, hardware addresses, unit conversions, etc. Each functionality class is represented by one and only one CAPFAST device schematic. Relevant device symbol parameters, such as instance macros and hierarchical port names, are automatically imported into the RDB from the CAPFAST device symbol files. Fig. 1 shows an entity relationship diagram of the relevant part of the ISAC device RDB.

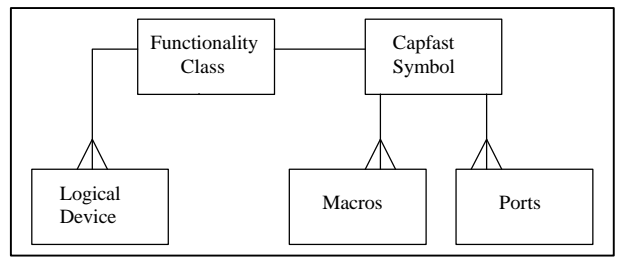

Figure 1: Partial entity relationship diagram of the ISAC device RDB

### 4.2 Automatic EPICS Database Generation

For the automatic generation of CAPFAST sub-system schematics, a tool *schgen* was implemented in Perl. This was greatly facilitated by the fact that both CAFAST schematic and symbol files are ASCII text files which can be easily interpreted and re-engineered. The schgen tool uses device information files, generated by the RDB, which contain all necessary device instance information such as device names, CAPFAST device symbol names, device instance macros and device instance hardware addresses. Additional information is extracted from CAPFAST device symbol files. Finally, a sub-system schematic is produced in CAPFAST format

The ISAC control system developer starts the generation of a sub-system IOC data from the Paradox RDB application main form. (S)he selects the desired ISAC sub-system as well as the subsystem I/O hardware type. Default values for the sub-system schematic name, scan parameters, etc can be overridden at this stage. The RDB then generates the device information files and spawns the schematic generation tool schgen, which produces the sub-system CAPFAST schematic file. An example of an automatically generated sub-system schematic viewed with CAPFAST is shown in Figure 2. This file is then translated into the sub-system IOC database by the standard CAPFAST/EPICS tools sch2edif and e2db.

## 5 SUMMARY

The move from interactive to automatic generation of CAPFAST sub-system schematics significantly reduced the number of errors in the ISAC runtime databases, mainly due to parameter checking in the underlying relational database application. The retaining of a unified graphical representation throughout the control system is well appreciated by all developers. The conversion of older sub-systems from interactive to RDB supported generation was recently completed.

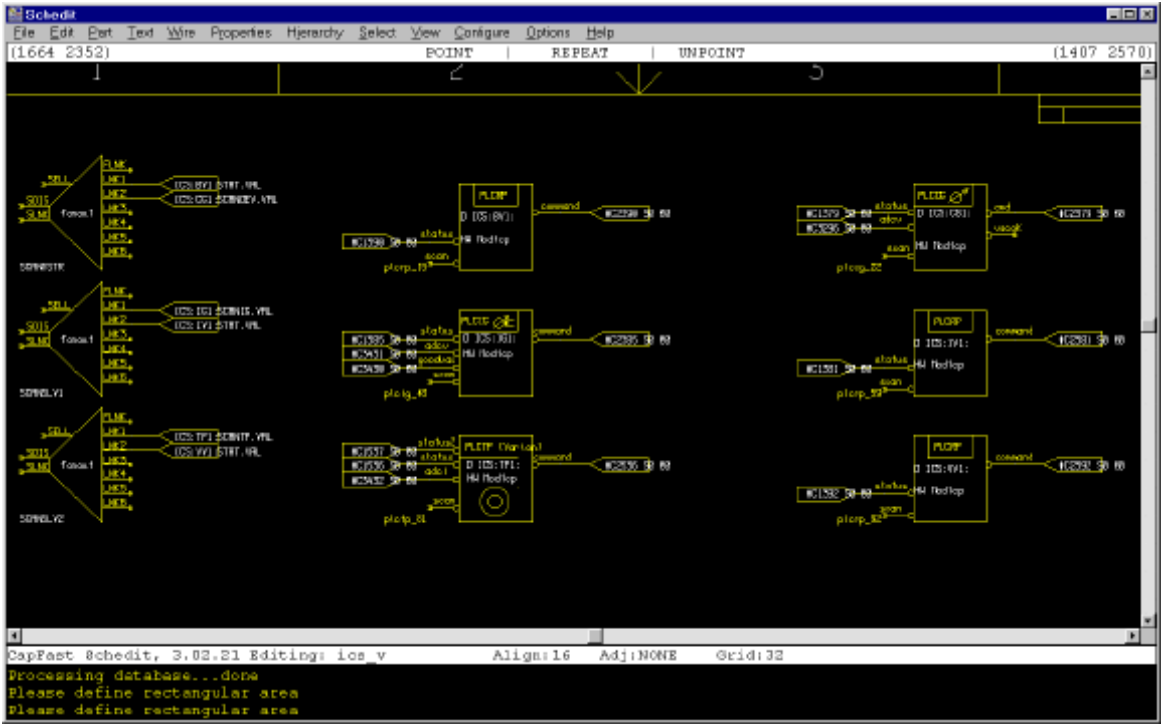

Figure 1: Example of an automatically generated sub-system schematic viewed with CAPFAST

## ACKNOWLEDGEMENTS

The author wants to thank J. Richards for helpful discussions about relational database design issues. Special thanks to E. Tikhomolov, who undertook the effort of integrating older sub-systems into the RDB with admirable diligence, and to R. Nussbaumer who provided IOC database comparison tools.